\documentstyle[12pt,psfig]{article}

\begin{document}
\baselineskip 0.7cm

\newcommand{\gsim}{ \mathop{}_{\textstyle \sim}^{\textstyle >} }
\newcommand{\lsim}{ \mathop{}_{\textstyle \sim}^{\textstyle <} }
\newcommand{\vev}[1]{ \left\langle {#1} \right\rangle }
\newcommand{\lsp}{ \left ( }
\newcommand{\rsp}{ \right ) }
\newcommand{\lmp}{ \left \{ }
\newcommand{\rmp}{ \right \} }
\newcommand{\llp}{ \left [ }
\newcommand{\rlp}{ \right ] }
\newcommand{\labs}{ \left | }
\newcommand{\rabs}{ \right | }
\newcommand{\EV} { {\rm eV} }
\newcommand{\KEV}{ {\rm keV} }
\newcommand{\MEV}{ {\rm MeV} }
\newcommand{\GEV}{ {\rm GeV} }
\newcommand{\TEV}{ {\rm TeV} }
\newcommand{\YR}{ {\rm yr} }
\newcommand{\mgut}{M_{GUT}}
\newcommand{\mint}{M_{I}}
\newcommand{\mgra}{M_{3/2}}
\newcommand{\mll}{m_{\tilde{l}L}^{2}}
\newcommand{\mdr}{m_{\tilde{d}R}^{2}}
\newcommand{\mllXX}[1]{m_{\tilde{l}L , {#1}}^{2}}
\newcommand{\mdrXX}[1]{m_{\tilde{d}R , {#1}}^{2}}
\newcommand{\mgy}{m_{G1}}
\newcommand{\mgl}{m_{G2}}
\newcommand{\mgc}{m_{G3}}
\newcommand{\nuR}{\nu_{R}}
\newcommand{\slL}{\tilde{l}_{L}}
\newcommand{\slLi}{\tilde{l}_{Li}}
\newcommand{\sdR}{\tilde{d}_{R}}
\newcommand{\sdRi}{\tilde{d}_{Ri}}
\newcommand{\e}{{\rm e}}
\newcommand{\bsub}{\begin{subequations}}
\newcommand{\esub}{\end{subequations}}
\newcommand{\btable}{\begin{table}[htbp]\begin{center}}
\newcommand{\etable}[1]{ \end{tabular}\caption{#1}\end{center}\end{table} }
\makeatletter
%
%
%
%
%
\newtoks\@stequation

\def\subequations{\refstepcounter{equation}%
  \edef\@savedequation{\the\c@equation}%
  \@stequation=\expandafter{\theequation}
  \edef\@savedtheequation{\the\@stequation}
  \edef\oldtheequation{\theequation}%
  \setcounter{equation}{0}%
  \def\theequation{\oldtheequation\alph{equation}}}

\def\endsubequations{%
  \ifnum\c@equation < 2 \@warning{Only \the\c@equation\space subequation
    used in equation \@savedequation}\fi
  \setcounter{equation}{\@savedequation}%
  \@stequation=\expandafter{\@savedtheequation}%
  \edef\theequation{\the\@stequation}%
  \global\@ignoretrue}


\def\eqnarray{\stepcounter{equation}\let\@currentlabel\theequation
\global\@eqnswtrue\m@th
\global\@eqcnt\z@\tabskip\@centering\let\\\@eqncr
$$\halign to\displaywidth\bgroup\@eqnsel\hskip\@centering
     $\displaystyle\tabskip\z@{##}$&\global\@eqcnt\@ne
      \hfil$\;{##}\;$\hfil
     &\global\@eqcnt\tw@ $\displaystyle\tabskip\z@{##}$\hfil
   \tabskip\@centering&\llap{##}\tabskip\z@\cr}

\makeatother


\begin{titlepage}

\begin{flushright}
UT-02-05
\end{flushright}

\vskip 0.35cm
\begin{center}
{\large \bf Neutrinoless Double Beta Decay with R-parity Violation}
\vskip 1.2cm
Yosuke Uehara

\vskip 0.4cm

{\it Department of Physics, University of Tokyo, Tokyo 113-0033 Japan}

\vskip 1.5cm

\abstract{We consider recently observed neutrinoless double beta decay 
in the context of the minimal supersymmetric standard model 
with R-parity violating couplings $\lambda^{'}$. We observe that most of
the current experimental bounds on the R-parity violating couplings
do not exclude the possibility that the neutrinoless double beta
decay is caused by R-parity violation. But if we consider $K-\bar{K}$ 
oscillation, we observe that we have to make the R-parity violating couplings
generation-dependent to accomodate with the observed 
neutrinoless double beta decay. And furthermore, we need some mechanism
to cancel the contribution to $K-\bar{K}$ mixing
from a large R-parity violating coupling. We realized this cancellation
by assuming that the first- and the second- generation of quark sector do not
couple with the first-generation lepton sector by R-parity violating
couplings except the term $W=\lambda_{111}^{'} L_{1} Q_{1} D_{1}^{c}$,
which is responsible for the observed neutrinoless double beta decay.}

\end{center}
\end{titlepage}

\renewcommand{\thefootnote}{\arabic{footnote}}
\setcounter{footnote}{0}

%
%
%
%

Recently evidence for neutrinoless double beta decay has been found by 
the HEIDELBERG-MOSCOW double beta decay experiment~\cite{DOUBLEBETA}. 
The half-life of $^{76}$Ge is reported to be:
\begin{eqnarray}
T_{1/2}^{0 \nu} &=& (0.8 - 18.3) \times 10^{25} {\rm yr}. 
\end{eqnarray}
This means that, lepton number is broken in nature.
In the Standard Model(SM), lepton number is conserved, and this evidence
becomes signature for physics beyond the SM.

We can realize lepton number violation in the R-parity violating
Minimal Supersymmetric Standard Model (MSSM) 
(for reviews, see \cite{REVIEW}). 
The R-parity violating couplings are:
\begin{eqnarray}
W=\lambda_{ijk} L_{i} L_{j} E_{k}^{c} + \lambda_{ijk}^{'} L_{i} Q_{j} D_{k}^{c} + \lambda_{ijk}^{''} U_{i}^{c} D_{j}^{c} D_{k}^{c} 
\label{RPV}
\end{eqnarray}
These terms violates lepton number and baryon number simultaneously,
and thus lead to rapid proton decay. So we must forbid 
some or all of these terms. Usually, to achieve that, 
a $Z_{2}$-symmetry called as ``R-parity'' is imposed. R-parity is defined as:
\begin{eqnarray}
R_{p} = (-1)^{3B+L+2S},
\end{eqnarray}
where B is the baryon number of the particle, L is the lepton number 
of the particle and S is the spin of the particle. If we impose
R-parity, all the couplings in equation (\ref{RPV}) are forbidden
, and no dangerous phenomena occur.

But there is another possibility. $Z_{3}$-symmetry is anomaly-free 
discrete gauge symmetry, and can protect proton from 
rapid decay \cite{DISCRETESYM}. This symmetry forbids 
baryon number violation, but allows lepton-number violation.
So it is worthwhile to investigate the lepton-number violating phenomena
as resultants of a $Z_{3}$-symmetry \cite{MYWORK}. 
The charge assignment of this $Z_{3}$-symmetry is shown in table \ref{Z3TAB}.
\btable
\begin{tabular}{|c|c|c|c|c|c|}
\hline
particle & $Q$ & $U^{c}$ & $D^{c}$ & $L$ & $E^{c}$ \\
\hline
charge & $1$ & $\alpha^{2}$ & $\alpha$ & $\alpha^{2}$ & $\alpha^{2}$ \\
\hline
\etable{Charge assignment under the discrete gauge
 symmetry. $\alpha^{3}=1$. \label{Z3TAB}}

Neutrinoless double beta decay was considered in the context of
the MSSM with lepton-number violating R-parity 
\cite{DOUBLEBETATHEORY1,DOUBLEBETATHEORY2}.
Detailed calculations for the neutrinoless double beta decay rate 
including nuclear matrix elements was done in~\cite{DOUBLEBETATHEORY2}. 
When only $\lambda^{'}$ couplings are considered, the Feynman diagrams
contributing to the neutrinoless double beta
decay are drawn in figure~\ref{DOUBLEBETAFIG}.
Since squark- and gluino-mediated process dominates, 
we drop the contribution from neutralino- and slepton-exchange 
diagrams~\cite{DOUBLEBETATHEORY2}.

Following reference~\cite{DOUBLEBETATHEORY2},
the recent result yields following constraints:
\begin{eqnarray}
1.6 \times 10^{-4} \left( \frac{m_{\tilde{q}}}{100 \GEV} \right)^{2} \left( \frac{m_{\tilde{g}}}{100 \GEV} \right)^{1/2} < \lambda_{111}^{'} < 3.6 \times 10^{-4} \left( \frac{m_{\tilde{q}}}{100 \GEV} \right)^{2} \left(\frac{m_{\tilde{g}}}{100 \GEV} \right)^{1/2}, \label{DOUBLEBETAEQ} \nonumber \\ 
\end{eqnarray}
where we assume $m_{\tilde{d}_{R}} = m_{\tilde{u}_{L}} \equiv m_{\tilde{q}}$.

By scanning the parameter region $100 \GEV < m_{\tilde{q}} < 2000 \GEV$,
$200 \GEV < m_{\tilde{g}} < 2000 \GEV$, we make 
a contour plot of the allowed values of $\lambda_{111}^{'}$. 
It is shown in figure \ref{CONTOURFIG}.
Here we have conservatively adopted $m_{\tilde{g}} > 200 \GEV$.
This figure shows the allowed region of $m_{\tilde{q}}$ 
and $m_{\tilde{g}}$ for given values of $\lambda^{'}$.
We can see that as $\lambda^{'}$ couplings become smaller,
the allowed region of $m_{\tilde{q}}$ and $m_{\tilde{g}}$ is
lowered. This is because if squark and gluino masses are heavy,
their contribution to the neutrinoless double beta decay becomes small.
 
It is interesting to compare the combined constraint on $\lambda^{'}$
v.s. squark and gluino masses obtained here, with those from other 
experimental results. There are many experimental 
results which can constrain $\lambda^{'}$. Hereafter, we study them in detail.

For example, the existence of R-parity violation leads to 
a violation of the universality of quark and lepton couplings to
the W boson. In the quark sector, the R-parity violating couplings 
$\lambda_{ijk}^{'} L_{i} Q_{j} D_{k}^{c}$ gives an additional
contribution to the quark semileptonic decay (e.g., in nuclear $\beta$ 
decay) like muon decay. The effective coupling becomes:
\begin{eqnarray}
\frac{g^{2}}{8 m_{W}^{2}} [ V_{ud} + r_{11k}^{'} (\tilde{d}_{R}^{k})],
\end{eqnarray}
where $r_{11k}^{'}$ is defined as:
\begin{eqnarray}
r_{11k}^{'}(\tilde{l}) = \frac{m_{W}^{2}}{g^{2}} \frac{|\lambda_{ijk}^{'}|^{2}}{m_{\tilde{l}}^{2}}.
\end{eqnarray}
The CKM matrix elements are experimentally determined from the ratio of
the $Q \rightarrow q e \nu_{e}$ to $\mu \rightarrow \nu_{\mu} e \nu_{e}$
partial widths. The experimental value is related to theoretical
quantities by
\begin{eqnarray}
|V_{ud}|_{\rm exp}^{2} &=& \frac{|V_{ud} + r_{11k}^{'} (\tilde{d}_{R}^{k})|^{2}}{|1 + r_{12k} (\tilde{e}_{R}^{k})|^{2}},
\end{eqnarray}
where $r_{12k}$ is defined like $r_{11k}^{'}$. 
A comparison with the experimental value:
\begin{eqnarray}
\sum_{j}|V_{u d_{j}}|_{\rm exp}^{2} &=& 0.9979 \pm 0.0021 \nonumber \\
\label{CCUNIVERSALITYEQ}
\end{eqnarray}
yields the limit \cite{BARGER-GIUDICE-HAN}:
\begin{eqnarray}
|\lambda_{11k}^{'}| < 0.03  \left( \frac{m_{\tilde{d}_{R}^{k}}}{100 \GEV} \right).
\label{CCUNIVERSALITYLIMITEQ}
\end{eqnarray}
at the $2 \sigma$ level.

This does not exclude the possibility that R-parity violation is
responsible for the neutrinoless double beta decay.
For example, for $m_{d_{R}}=m_{\tilde{q}}=500 \GEV$, this limit
becomes $|\lambda_{11k}^{'}| < 0.15$, which is compatible with
the allowed values of $\lambda_{111}^{'}$ 
shown in figure \ref{CONTOURFIG}.

The decay rate of pion into electron and muon is also changed
in the presence of the R-parity violating couplings. The ratio 
$R_{\pi} \equiv \Gamma(\pi \rightarrow e \nu)/\Gamma(\pi \rightarrow \mu
\nu)$ is
\begin{eqnarray}
\frac{R_{\pi}({\rm expt})}{R_{\pi}({\rm SM})} = 0.991 \pm 0.18.
\label{EMUTAUUNIVERSALITYEQ}
\end{eqnarray}
R-parity violation gives an effective contribution to $R_{\pi}$
\cite{BARGER-GIUDICE-HAN}:
\begin{eqnarray}
R_{\pi} = R_{\pi} (SM) \left[ 1 + \frac{2}{V_{ud}} [r_{11k}^{'} (\tilde{d}_{R}^{k}) -r_{21k}^{'} (\tilde{d}_{R}^{k})] \right].
\end{eqnarray}
The experimental value (\ref{EMUTAUUNIVERSALITYEQ}) set upper limit on 
the R-parity violating couplings as:
\begin{eqnarray}
|\lambda_{11k}^{'}|<0.05 \left( \frac{m_{\tilde{d}_{R}^{k}}}{100 \GEV} \right).
\end{eqnarray}
This is weaker limit compared to the equation (\ref{CCUNIVERSALITYLIMITEQ}),
thus we can neglect this limit in this study.

The decay $K^{+} \rightarrow \pi \nu \bar{\nu}$ is also modified
in the presence of the R-parity violating couplings \cite{KDECAY}.
We obtain:
\begin{eqnarray}
\frac{\Gamma[K^{+} \rightarrow \pi^{+} \nu_{i} \bar{\nu}_{i}]}{\Gamma[K^{+} \rightarrow \pi^{0} \nu e^{+}]} = \left( \frac{|\lambda_{ijk}^{'}|^{2}}{4 G_{F} m_{\tilde{d}_{R}^{k}}^{2}} \right)^{2} \left( \frac{|V_{j1} V_{j2}^{*}|}{|V_{12}^{*}|} \right)^{2}.
\end{eqnarray}
So using $B(K^{+} \rightarrow \pi^{+} \nu \bar{\nu}) \lsim 4.4 \times
10^{-10}$ \cite{KPINUNU} and $B(K^{+} \rightarrow \pi^{0} \nu
e^{+})=0.0482$ \cite{KPINUE}, we obtain the constraint \cite{KDECAY,KPINUNU}.
\begin{eqnarray}
|\lambda_{ijk}^{'}| < 0.0056 \left( \frac{m_{\tilde{d}_{R}^{k}}}{100 \GEV} \right).
\end{eqnarray}
This constraint is stronger. For example, take $m_{\tilde{d}_{R}^{k}}=900
\GEV$, then $\lambda_{111}^{'}<0.05$. From the figure \ref{CONTOURFIG}
we can see that gluino mass is constrained in the region:
\begin{eqnarray}
m_{\tilde{g}} \lsim 1100 \GEV.
\end{eqnarray}

Other experiments, like $K-\bar{K}$ oscillation, 
and $B-\bar{B}$ oscillation give stronger limits 
on the lepton number violating couplings \cite{KKBB}.
But their limit always contain products of two $\lambda^{'}$.
Thus we cannot state strongly that we can derive upper limit
on $\lambda_{111}^{'}$. For example, $K-\bar{K}$ 
oscillation gives \cite{KKBB}:
\begin{eqnarray}
{\rm Re} \left[ \sum_{i,j,j^{'}} \left( \frac{100 \GEV}{m_{\tilde{\nu}}} \right)^{2} \lambda_{ij2}^{'*} \lambda_{ij^{'}1}^{'} V_{j1}^{*} V_{j^{'}2} \right] < 4.5 \times 10 ^{-9}. \label{KKmixingeq}
\end{eqnarray}
So we cannot extract the information of $\lambda_{111}^{'}$ from 
$K-\bar{K}$ oscillation. Of course if we assume generation-independence
of the $\lambda^{'}$ couplings, we can estimate:
\begin{eqnarray}
\lambda_{111}^{'} \lsim 10^{-4} \left( \frac{m_{\tilde{\nu}}}{100 \GEV} \right).
\end{eqnarray}
As we can see from figure \ref{CONTOURFIG}, 
this is so strong constraint that we cannot explain the 
observed neutrinoless double beta decay if we impose this constraint.
So we can say that if the obserbed neutrinoless double beta decay
is truly the result of R-parity violation, the $\lambda^{'}$ couplings
are not generation-independent.

But there still exists a non-trivial problem that how such a large
$\lambda_{111}^{'}$ coupling can be consistent with the stringent
bound from $K-\bar{K}$ mixing (equation (\ref{KKmixingeq})). 
We should make such a large coupling be cancelled by some mechanism
to accomodate with the stringent bound from $K-\bar{K}$ mixing. 
One way is to assume that
\begin{eqnarray}
\lambda_{112}^{'} = \lambda_{121}^{'} = \lambda_{122}^{'} = 0.
\end{eqnarray}
In this case, the contribution from $\lambda_{111}^{'}$
to the equation (\ref{KKmixingeq}) becomes
\begin{eqnarray}
{\rm Re} [\lambda_{111}^{'} \lambda_{132}^{'*} V_{31}^{*} V_{12}] < 4.5 \times 10^{-9}. \label{KKmixingeq2}
\end{eqnarray}
We substitute $|V_{31}| \sim 0.003$, $|V_{12}| \sim 0.22$ and 
$\lambda_{111}^{'} \sim 0.005$ into (\ref{KKmixingeq2}), then
we get
\begin{eqnarray}
\lambda_{132}^{'*} < 1.4 \times 10^{-3},
\end{eqnarray}
which is moderate value compared to $\lambda_{111}^{'} = 5 \times 10^{-3}$
. So we conclude that the large value of $\lambda_{111}^{'}$
can be consistent with the stringent bound from $K-\bar{K}$ mixing.

To summarize, we consider the neutrinoless double beta decay
in the context of the Minimal Supersymmetric Standard Model
with the lepton-number violating R-parity couplings. We observe
that most of the current experiments do not exclude the possibility
that R-parity violation is the source of the observed neutrinoless
double beta decay. But if the R-parity violating couplings are
generation-independent, the constraint on $K-\bar{K}$ oscillation
excludes this possibility. Generation-dependency and some mechanism
to cancel a large $\lambda_{111}^{'}$ coupling contribution to
$K-\bar{K}$ oscillation is needed. We realized this cancellation by assuming 
$\lambda_{112}^{'}=\lambda_{121}^{'}=\lambda_{122}^{'}=0$, 
namely the first- and the second-generation
of quark sector do not couple with the first-generation lepton sector
by the R-parity violating couplings, except the coupling which is
responsible for the neutrinoless double beta decay, $\lambda_{111}^{'}$. 

{\bf Note Added}

After the submittion of this letter, we learned from Dr.Liu that
they considered within the framework of R-parity violating
supersymmetry, the sleptons play a partial role in electroweak symmetry
breaking.  The scalar neutrinos get non-vanishing vacuum expectation values
(vevs).  These non-zero vevs break the family symmetry (say, $Z_3$) naturally.
This breaking of the family symmetry may result in the 
realistic pattern of the fermion masses \cite{DU-LIU1,DU-LIU2}.
To be specific, Refs. \cite{DU-LIU1,DU-LIU2} 
proposed that the muon mass originates from
the sneutrino vevs, whereas tau from Higgs.
Neutrino masses are discussed in Ref. \cite{LIU}.
Especially in \cite{LIU}, they have obtained 
an electron-neutrino Majorana mass to be around 0.1 eV.

And we also learned from Dr. Dedes that the bounds on all R-parity
violating couplings have been collated and updated in their paper
\cite{ALLANACH-DEDES-DREINER}.

\noindent
{\bf Acknowledgment}

We thank F.~Borzumati and T.~Yanagida for stimulating discussions.
In particular, much of this work was done in the collaboration with
F.~Borzumati. Y.U. thank Japan Society for the Promotion of Science
for finantial support.

\begin{figure}[htbp]
\centerline{\psfig{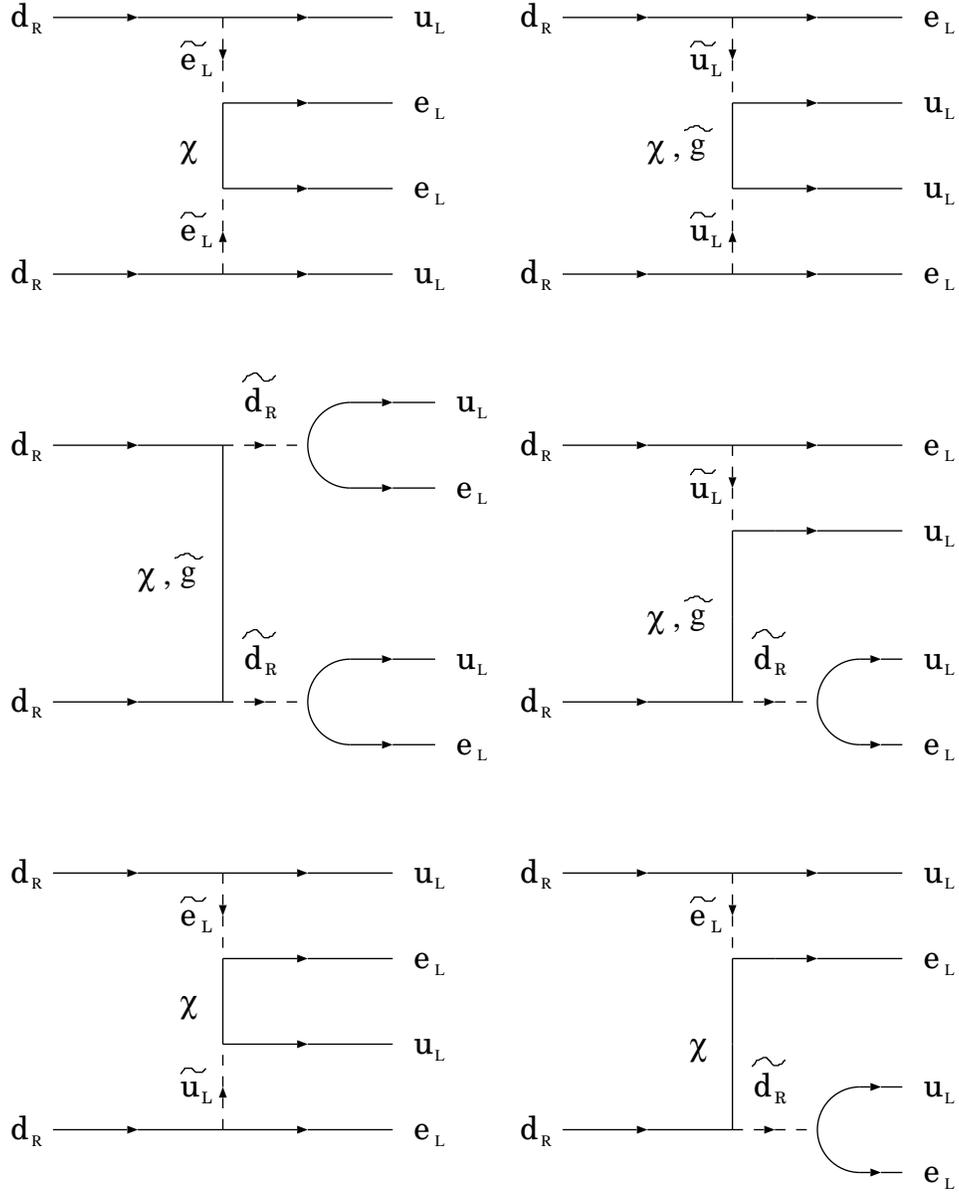}}
\caption{The processes relevant for neutrinoless double beta decay.}
\label{DOUBLEBETAFIG}
\end{figure}

\begin{figure}[htbp]
\centerline{\psfig{file=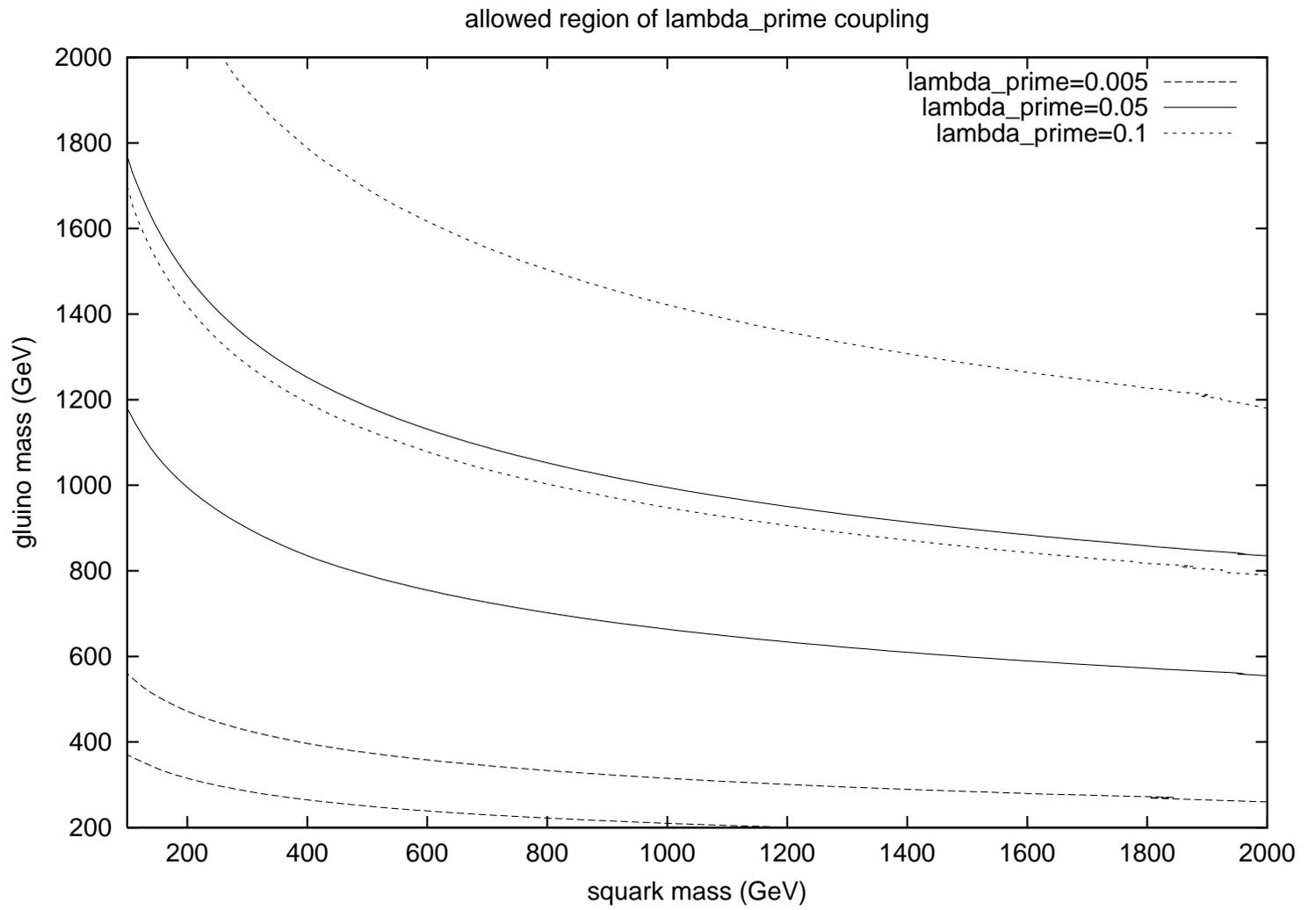,height=14cm}}
\caption{The allowed region for given values of $\lambda_{111}^{'}$.
Here we take $\lambda_{111}^{'}=0.005, \ 0.05, \ 0.1$.}
\label{CONTOURFIG}
\end{figure}

\end{document}